\documentclass[manuscript]{aastex}

\slugcomment{070125-070315}

 \shorttitle{On the Interpretation  for Low-High Frequency QPO}
 \shortauthors{C. M.  ZHANG et al.}

\def\lsim{\lower.5ex\hbox{$\; \buildrel < \over \sim \;$}}
\def\gsim{\lower.5ex\hbox{$\; \buildrel > \over \sim \;$}}
\def\lax {\ifmmode{_<\atop^{\sim}}\else{${_<\atop^{\sim}}$}\fi}
\def\gax {\ifmmode{_>\atop^{\sim}}\else{${_>\atop^{\sim}}$}\fi}
\def\etal{{\it et al.\/} }
\def\gtorder{\mathrel{\raise.3ex\hbox{$>$}\mkern-14mu
\lower0.6ex\hbox{$\sim$}}}
\def\ltorder{\mathrel{\raise.3ex\hbox{$<$}\mkern-14mu
\lower0.6ex\hbox{$\sim$}}}

\def\pmb#1{\setbox0=\hbox{#1}%
\kern-0.015em\copy0\kern-\wd0 \kern0.03em\copy0\kern-\wd0
\kern-0.015em\raise0.0433em\box0 }

\def\be{\begin{equation}}
\def\ee{\end{equation}}
\def\bea{\begin{eqnarray}}
\def\eea{\end{eqnarray}}

\def\lan{ \langle}
\def\ran{ \rangle}
\def\ov{ \over}

\def\mdot{\ifmmode \dot M \else $\dot M$\fi}    
\def\mxd{\ifmmode \dot {M}_{x} \else $\dot {M}_{x}$\fi}
\def\med{\ifmmode \dot {M}_{Edd} \else $\dot {M}_{Edd}$\fi}
\def\bff{\ifmmode B_{f} \else $B_{f}$\fi}

\def\apj{\ifmmode ApJ \else ApJ \fi}    
\def\apjl{\ifmmode  ApJ \else ApJ \fi}    %
\def\aap{\ifmmode A\&A \else A\&A\fi}    %
\def\mnras{\ifmmode MNRAS \else MNRAS \fi}    %
\def\nat{\ifmmode Nature \else Nature \fi}
\def\prl{\ifmmode Phys. Rev. Lett. \else Phys. Rev. Lett.\fi}
\def\prd{\ifmmode Phys. Rev. D. \else Phys. Rev. D.\fi}

\def\mdot{\ifmmode \dot M \else $\dot M$\fi}    
\def\mxd{\ifmmode \dot {M}_{x} \else $\dot {M}_{x}$\fi}
\def\med{\ifmmode \dot {M}_{Edd} \else $\dot {M}_{Edd}$\fi}
\def\bff{\ifmmode B_{f} \else $B_{f}$\fi}

\def\apj{\ifmmode ApJ \else ApJ \fi}    
\def\apjl{\ifmmode  ApJ \else ApJ \fi}    %
\def\aap{\ifmmode A\&A \else A\&A\fi}    %
\def\mnras{\ifmmode MNRAS \else MNRAS \fi}    %
\def\nat{\ifmmode Nature \else Nature \fi}
\def\prl{\ifmmode Phys. Rev. Lett. \else Phys. Rev. Lett.\fi}
\def\prd{\ifmmode Phys. Rev. D. \else Phys. Rev. D.\fi}

\def\mwd{ {\rm M_{WD}} }
\def\mns{ {\rm M_{NS}} }
\def\rwd{ {\rm R_{WD}} }
\def\rns{ {\rm R_{NS}} }
\def\awd{ {\rm A_{WD}} }
\def\ans{ {\rm A_{NS}} }
\def\xwd{ {\rm X_{WD}} }
\def\xns{ {\rm X_{NS}} }

\def\ms{ {\rm M_{\odot}} }    
\def\na{\ifmmode \nu_{A} \else $\nu_{A}$\fi}
\def\nk{\ifmmode \nu_{K} \else $\nu_{K}$\fi}
\def\nma{\ifmmode \nu_{\rm MA} \else $\nu_{\rm MA}$\fi}    %
\def\ns{\ifmmode \nu_{{\rm s}} \else $\nu_{{\rm s}}$\fi}
\def\no{\ifmmode \nu_{1} \else $\nu_{1}$\fi}    
\def\nt{\ifmmode \nu_{2} \else $\nu_{2}$\fi}
\def\nhigh{\ifmmode \nu_{\rm high} \else $\nu_{\rm high}$\fi}
\def\nlow{\ifmmode \nu_{\rm low} \else $\nu_{\rm low}$\fi}    
\def\ntk{\ifmmode \nu_{2k} \else $\nu_{2k}$\fi}    
\def\dnmax{\ifmmode \Delta \nu_{max} \else $\Delta \nu_{2max}$\fi}
\def\ntmax{\ifmmode \nu_{2max} \else $\nu_{2max}$\fi}    
\def\nomax{\ifmmode \nu_{1max} \else $\nu_{1max}$\fi}    
\def\nh{\ifmmode \nu_{\rm HBO} \else $\nu_{\rm HBO}$\fi}    
\def\nqpo{\ifmmode \nu_{QPO} \else $\nu_{QPO}$\fi}    
\def\nz{\ifmmode \nu_{o} \else $\nu_{o}$\fi}    
\def\nht{\ifmmode \nu_{H2} \else $\nu_{H2}$\fi}    
\def\ns{\ifmmode \nu_{s} \else $\nu_{s}$\fi}    %
\def\nb{\ifmmode \nu_{{\rm burst}} \else $\nu_{{\rm burst}}$\fi}
\def\nkm{\ifmmode \nu_{km} \else $\nu_{km}$\fi}    %
\def\ka{\ifmmode \kappa \else \kappa\fi}    %
\def\dn{\ifmmode \Delta\nu \else \Delta\nu\fi}    %

\begin{document}

\title{The Interpretations  For  the Low  and High Frequency
 QPO Correlations  of X-ray Sources Among  White Dwarfs, Neutron Stars and Black Holes}

\author{C. M. Zhang,  H. X. Yin,   Y. H. Zhao}
\affil{National Astronomical Observatories, Chinese Academy of
Sciences, Beijing 100012, China, zhangcm@bao.ac.cn }


\begin{abstract}
{\bf
It is found that   there exists an empirical linear relation
 between the high frequency $\nhigh$ and low  frequency $\nlow$
 of  quasi-periodic oscillations (QPOs)  for black hole candidate (BHC),
 neutron star (NS)  and  white dwarf (WD) in the binary systems, which
 spans five orders of magnitude  in frequency.
 For the NS Z (Atoll) sources,
  $\nu_{high}$ and $\nu_{low}$  are identified as  the  lower kHz QPO
  frequency and   horizontal branch oscillations (HBOs) $\nh$
  (broad noise components); for the black hole candidates and low-luminosity
 neutron stars, they are the  QPOs and    broad noise  components
at frequencies between 1 and 10 Hz; for WDs, they are the ``dwarf nova
oscillations'' (DNOs) and QPOs   of cataclysmic variables (CVs).
 To interpret this relation, our model ascribes $\nu_{high}$
 to the Alfv\'en wave oscillation frequency  at a preferred  radius
 and $\nu_{low}$  to  the same mechanism at another radius.
  Then, we  can obtain   $\nlow = 0.08 ~\nhigh$ and   the  relation between
  the upper kHz QPO frequency $\nt$ and  HBO  to be  $\nh \simeq ~56~ ({\rm
Hz})~ (\nt/{\rm kHz})^{2}$, which are in accordance with the
observed empirical relations.  Furthermore, some implications of model are
discussed, including  why  QPO frequencies of white dwarfs and
neutron stars  span  five orders of magnitude in frequency.
}
\end{abstract}

 \keywords{accretion --- stars: neutron--- white dwarfs ---
 stars: oscillations  --- X-ray: binaries }

   \maketitle

\section{Introduction}

To  search the timing  properties of the aperiodic variability of accreting
X-ray binaries, the Rossi X-ray Timing Explorer ({\em RXTE}) was
launched in the end of 1995,  and  has risen  many discoveries
(e.g., van der Klis 2000, 2005, 2006 for recent reviews).  The quasi
periodic oscillations (QPOs) and broad noise components are observed
in the power spectra of neutron stars (NSs) and black hole
candidates (BHCs) in low-mass X-ray binaries (LMXBs),  which covers
many QPO frequency bands, such as the kilohertz (kHz) QPO    $\simeq
200 - 1300$ Hz that has been found  in about 28 LMXBs (e.g., van der
Klis 2006) and typically occurs in pairs: the  upper kHz QPO
$\nt$ and lower kHz QPO $\no$,  the $\simeq 15 - 70$ Hz horizontal
branch oscillations (HBOs) $\nh$ in  Z sources  or broad noise
components in Atoll sources (on the definition of Atoll and Z, see
 Hasinger \& van der Klis 1989; Hasinger 1990).
 The numerous QPOs and {\bf broad noise components} in the 0.1 - 1300~Hz
  range  are found to follow the tight correlations covering
three orders of magnitude in frequency (see, e.g., Psaltis, Belloni
\& van der Klis  1999, hereafter PBK99;  Belloni, Psaltis \& van der
Klis 2002, hereafter BPK02).
For the  bright  Z sources, the  two QPO features are the HBO and
the lower kHz QPO, respectively. For the BHCs and low-luminosity
  sources, they are the low-frequency QPO and a broad noise component
at frequencies between 1 and 10 Hz.   PBK99 and BPK02 have
demonstrated that these NS and BHC low and high frequencies follow a
tight correlation over 3 orders of magnitude in frequency;  the
correlation can be {\bf approximately described by a linear relation as }
 $\nlow = 0.08 \nhigh$ (Titarchuk \& Wood 2002, hereafter TW02, see also
figure 1).

Moreover, dwarf nova oscillations (DNOs) and quasi-periodic oscillations
(QPOs) were discovered a few decades ago (Warner \& Robinson 1972;
Patterson et al. 1977; Warner 2004 for a review).  Motivated by
Warner \& Woudt (2002ab), Mauche (2002) studies  the data of  the
{\it Extreme Ultraviolet Explorer\/} deep survey photometer and  the
{\it Chandra X-ray Observatory\/} Low Energy Transmission Grating
Spectrograph to investigate the extreme-ultraviolet (EUV) and soft
X-ray oscillations  of the dwarf nova SS~Cyg in outburst.
 To combine with the optical data of Woudt \&  Warner (2002) for
VW~Hyi,  he  extends  the $\nlow$-$\nu_{\rm high}$   correlation
($\nu_{\rm  low} \approx 0.08\, \nu_{\rm high}$) for NS/BHC  LMXBs
nearly two orders of magnitude in frequency.
 This correlation identifies the high-frequency quasi-coherent
 oscillations (so-called ``dwarf nova
oscillations'', DNOs) of cataclysmic variables (CVs) with the lower
kHz  QPOs of LMXBs, and the low-frequency QPOs of CVs with the HBOs
   of LMXBs.
 %

 Furthermore,  systematically studied and improved by Warner et al (2003)
and Warner \& Woudt (2004ab), the 27 white dwarf   CVs and LMXBs are found to
  share a common correlation in the QPO frequencies  over nearly five
  orders of magnitude in frequency, which also establishes  the connection
   between the CV and LMXB QPOs.
In addition to the  frequencies, the DNOs of CVs and  lower  kHz
QPOs of NS LMXBs  are similar in that they have relatively high
coherence and high amplitudes,  and  sometimes occur in pairs
 (Warner et al 2003;  Warner \& Woudt 2004ab; 2006; Pretorius et al. 2006).
However,  while the similar QPO phenomena occur in CVs and BH LMXBs,
 they are not totally identical (e.g. Kluzniak et al. 2005).


On the origins  for QPO productions,  some  models exist for LMXBs.
 The beat frequency models (e.g.,  Miller, Lamb \& Psaltis 1998;
 Strohmayer et al.\ 1996;  Alpar \& Shaham 1985) take
  the NS spin-frequency  as  one of the
characteristic frequencies, {\bf however it is commented recently that
 this  model still  confronts the  severe  arguments or modifications
 in interpreting the varied twin kHz QPO separations and the correlation
 between the upper and lower kHz QPO frequencies
 (see, e.g.,  Swank 2004;  Belloni et al. 2005, 2007; van der Klis 2006;
 Zhang et al. 2006}).
 In the relativistic precession models
(see, e.g., Stella \& Vietri 1999; Stella et al. 1999; Morsink
2000), all QPO frequencies arise from general relativistic
frequencies in the  accretion disk, i.e., the upper
 and lower kHz QPOs ($\nt$ and $\no$) are
 ascribed to the orbital Keplerian frequency $\nk$ and the
periastron precession of it at the same orbit, respectively, whereas
  the HBO frequency   is ascribed to the nodal
 frequency of the tilt disk by the gravitomagnetism precession,
 frame-dragging  effect of Einstein's general relativity
 or the Lense-Thirring effect (Stella \& Vietri 1998; Stella et al. 1999).
  Although  the  empirical quadratic  relation between the
  HBO and  upper kHz QPO
can be derived in the relativistic precession model,
it needs the high spin frequency,
such as {\bf $\sim 900$}  Hz in  central object, and large moment of
inertia (see, e.g.,  Psaltis et al 1999), which are inadequate for
WD sources because of their  weak gravitational field  and slow
rotation (see, e.g., Mauche 2002; TW02; Warner \& Woudt 2004ab).

%

 More promising are models that interpret
QPOs as manifestations of disk accretion onto any low-magnetic field
compact objects, such as the transition layer model of Titarchuk and
collaborators (Titarchuk, Lapidus \& Muslimov  1998; Titarchuk et al
1999;  Titarchuk \& Osherovich 1999, 2001; Titarchuk et al 2001;
TW02).
 In the model of TW02   the high frequency
 $\nhigh$ (identified as  the lower kHz QPO frequency $\no$)  is ascribed
to the Keplerian  frequency $\nk$ at the outer (adjustment) radius
and the low frequency $\nlow$ (identified as the HBO) is ascribed
to the magnetoacoustic (MA) oscillation frequency $\nma$,
 and they demonstrate that the observed correlation of
 the low-high frequency  is perfectly described by the dependence of
 the inferred MA frequency $\nma$ on the Keplerian frequency $\nk$.
 Moreover, as for the QPO models that are nothing to do with the
presence or absence of the hard surface of the compact object,
 there exist the  mode oscillation models   in the accretion disc
 (see, e.g., Li \& Narayan 2004;  Rezzolla, et al 2003ab;
 Kato 2001; Nowak et al 1997; Wagoner 2001, references
 therein) and the orbital resonance model (Abramowicz et al. 2001, 2002, 2003;
 Rebusco 2004; Petri 2005; Horak \& Karas 2006), which needs the further
 refinements  if  applied to interpret the low and  high frequency correlations.

In this paper,  based on the Alfv\'en wave oscillation model that
has successfully  explained the  kHz QPOs of NS LMXBs (Zhang 2004),
 we seek to offer a unified prescription to  explain
 the similar  correlation between
the low and high frequencies found in CVs and LMXBs. In \S 2,
 our model formulation   is described, and
 the derivation of the low-high frequency correlation
 is   presented as well. The consequences and conclusions  are
 summarized in \S 3.

\section{Description of the model and its applications}

    In the model by Zhang (2004),  it ascribes
the upper and lower kHz QPO frequencies ($\nt$ and $\no$) to the
Keplerian frequency and the MHD Alfv\'en wave oscillation frequency
at a special  radius r,  where
 \be \nt(X) =
1850 {}A{}X^{3/2} ~({\rm Hz})= 1295 (A/0.7) {}X^{3/2} ~({\rm Hz})\,,
\label{nt} \ee with  X=R/r,
 ${\rm A = (m/R_{6}^{3})^{1/2}}$, the scaled stellar radius
  ${\rm R_{6}=R/10^{6}cm}$ and  mass in the unit of solar mass ${\rm m={M/\ms}}$,

 \be \no (X) =
 \nt(X){} X (1 - \sqrt{1-X})^{1/2}  = 1295{} (A/0.7) {}F(X)\,,
 \label{no}
\ee \be F(X) = X^{3}/(1 + \sqrt{1-X})^{1/2}\,. \label{fx} \ee
 {\bf The inferred averaged value of the parameter X
 $\lan X \ran\simeq 0.88$ comes from the
best fit of the kHz QPOs of  four detected  typical kHz QPO    sources
and it is not an optimization, then from $\lan X \ran\simeq 0.88$
it infers that  the position of producing the  kHz QPO is  close to NS
surface,  about $\sim 10\%$ of star radius  away (Zhang 2004).
 }
 It is remarked
 that our  model  (Zhang 2004), successfully applied in
interpreting the  kHz QPO correlations of NS LMXBs,  just needs the
accretion flow in the disk around a gravitational source with the
dipole  magnetic moment inside an innermost boundary,   and  it does
not necessarily  require the presence or absence of a hard surface
of compact object.

 {\bf To study the mechanisms of the low and high frequencies detected in
 NS/BH/WD systems by means of our kHz QPO model (Zhang 2004),
 we ascribe  the high frequency $\nhigh$ to
 the  Alfv\'en wave oscillation frequency  at a preferred radius  r
 and the low frequency $\nlow$ to be the same mechanism occurred
  at another radius of about $\phi$r, where $\phi$ is a parameter determined
  by the observation.
  Therefore, these low and high  frequencies ($\nlow$  and $\nhigh$)  are
conveniently arranged as follows
 \be
\nhigh = \no (X)\,, \label{nhigh}
 \ee
 \be
\nlow =  \no (\phi^{-1} X)\,,
\label{nlow}
 \ee
  and then we have the
ratio between the low and high frequencies,
 \be
{\nlow \ov \nhigh} = K(\phi, X)\,, \label{lh} \ee where ratio quantity
 $K(\phi, X)$ is defined as \be
 K(\phi, X) = 0.094 ({\phi \ov 2.2})^{-3} g(\phi, X)\,,
 \label{kx}
\ee \be g(\phi, X) =  [ \frac{1+\sqrt{1-X}}{1+\sqrt{1-\phi^{-1}
X}}]^{1/2} \le 1\,.
 \label{gx}
\ee
 Because  the position parameter  X=R/r is in the range of
   $0 \le X \le 1$ from its definition,
 we find that   $K(\phi, X) $ is mainly determined  by $\phi$ but it is
 insensitive  to  X,  and even more important
  the ratio quantity $K(\phi, X)$ is independent of the parameter
 A, where $A^{2} = m/R_{6}^{3}$ represents
  a measurement of the  average  mass density of compact  object.
 }%
  Therefore Eq.(\ref{lh}) can
be applicable in any compact objects, from WDs to NSs.
 For  the possible values of $\phi = 2.2$ and X=0.88,   from
 Eq.(\ref{kx}) and Eq.(\ref{gx}) we obtain  the ratio
  $K\simeq0.08$, which satisfies the need of empirical relation by TW02.

The correlation  $\nlow$ versus  $\nhigh$ is plotted in figure 1,
together with the  measured  data of CVs and LMXB sources  from
BPK02, Mauche (2002)  and Warner \& Woudt (2004), respectively. The
three fitted lines, from top to bottom, represent $\nlow/\nhigh =
0.11, 0.08, 0.06$,  which can be obtained by setting
 the parameter  values of  $\phi=2.1, 2.2, 2.3$ and X=0.88,
 respectively.  It is noticed   that the
  agreement  between the model and the observed QPO data is quite
 good for the convenient choice of  parameters X and $\phi$.
 It is stressed  that  the ratio  $K(\phi, X)$ in   Eq.(\ref{kx})
  has nothing to do with the parameter A  and  is almost  independent
  of the parameter X.  Therefore,  the $\nlow$ versus $\nhigh$ plot
   in figure 1  will represent the properties of
 any compact objects (CVs and LMXBs) that  reflect
a  common feature of the accretion flow around the gravitational
 sources and  have  no direct  correlations to the mass, radius,
 hard surface, magnetic field and  spin frequency of the specific
 source.

  If we assume the  position parameter X to be
similar  for the QPO sources of WDs and  NSs, then
 for the typical  choice of  mass and radius  of WD and NS
(e.g.,  Shapiro \& Teukolsky 1983), such as
  $\mwd=1.0\ms$,  $\rwd=5\times10^{8}$ cm,   $\mns=1.4\ms$
  and $\rns=1.5\times10^{6}$ cm, respectively,  we can obtain,
\be {\nhigh({\rm WD}) \ov \nhigh({\rm NS})} = {\awd \ov \ans}
{F(\xwd) \ov F(\xns)} = ({\mwd \ov \mns})^{1/2} ({\rwd \ov
\rns})^{-3/2} {F(\xwd) \ov F(\xns)} \sim 10^{-4}\,, \label{wdns} \ee
where we have set the position parameter  $\xwd \sim \xns$ or
${F(\xwd) \ov F(\xns)} \sim 1$ (see Eq. \ref{fx}).
 If the WD mass (radius) is lower (higher) than the typical
 value  or  the condition $\xwd > \xns$ happens,
 then the above ratio in  Eq.(\ref{wdns})
  can be  even as low as $\sim 10^{-5}$,
which may interpret the extended low-high frequency correlations in
CVs and LMXB  systems in five orders of magnitude  in frequency, as
shown in figure 1.  In fact, there are two parameters to modulate
the QPO frequency  in Eq.(\ref{no}) or Eq.(\ref{nhigh}),
 A and X, representing the compactness  of source and
 the position to exhibit QPO respectively.
Therefore for the same source (A is same),
 the parameter X will modulate  the QPO frequency,
 while for the different sources  the
parameter A will modulate  the QPO frequency. However, the fact
that the ratio of low frequency to high frequency is independent
of parameter A and weakly depends on the parameter X can interpret
the QPO similarities in different sources and in different
accretion rates of the same source (see figure 1).

 As an assumption that $\nlow$ is identified as the
 HBO frequency $\nh$
 for Z sources (or Broad band noises for Atoll sources),
 i.e., $\nh=\nlow$, we obtain the theoretical correlation
between the HBO frequency $\nh$ and the upper kHz QPO frequency
$\nt$ from Eqs.(\ref{nt}) and (\ref{nlow}) as,
 \be
{\nh} = 74 {}\; ({\rm Hz}) {}\; ({\phi \ov 2.2})^{-3} (A/0.7)^{-1}
 (\nt/{\rm kHz})^{2} f(\phi X) \,,
 \label{nhnt}
\ee \be f(\phi X) =  1/(1+\sqrt{1 - X/\phi})^{1/2}\,.
 \label{fpx}
\ee After substituting   $\phi = 2.2$, the averaged values $\lan X
\ran $=0.88 and A = 0.7 inferred from the typical kHz QPO sources
(Zhang 2004), into Eq.(\ref{nhnt}),
 we obtain  
 the simplified  theoretical correlation between $\nh$ and $\nt$ as
\be \nh \simeq ~56~ ({\rm Hz})~ (\nt/{\rm kHz})^{2}\;,
 \label{nhnt2}
\ee
 which is consistent with the observed empirical relation
(see, e.g., Psaltis et al 1999; Swank 2004; van der Klis 2000, 2006).


\section{Consequences and Discussions}

In this  paper, a  unified interpretation of QPO low-high frequency
 correlations observed in CVs and LMXB  sources has been proposed
 by means of  the  Alfv\'en wave oscillation model for kHz QPOs (Zhang 2004),
  where we identify the high frequency $\nhigh$ to be the lower
  kHz QPO frequency $\no$ and the low frequency $\nlow$ to be the same
mechanism of  $\no$ at the different radius. While,
   the linear relation between  $\nlow$ and $\nhigh$ can be
  obtained as $\nlow = 0.08 \nhigh$,
 which is consistent with the detected empirical
   relation  (see, e.g., TW02; van der Klis 2000, 2006;
 Psaltis et al 1999;  PBK99; BPK02),
    firstly  established by PBK99 and  BPK02
  for  LMXBs   and subsequently extended by Warner \& Woudt  (2002ab),
  Mauche (2002),  Warner et al (2003)
  and Warner \& Woudt  (2004ab) for WD   binaries.
  In addition,   the quadratic  correlation between
 the low frequency $\nlow$ (or HBO $\nh$) and
 upper kHz QPO frequency $\nt$ is approximately obtained as
 $\nh (\nlow) \simeq ~56~ ({\rm Hz})~ (\nt/{\rm kHz})^{2}$, which
 is also  in accordance with the observed empirical relation in
 LMXBs
(see, e.g, van der Klis 2000, 2006; Psaltis et al 1999; Stella et al
1999).

 Undoubtedly,
 these studies suggest that the same types of variabilities
occurred in  CVs and LMXBs  can severely constrain
 the theoretical models  (see, e.g., TW02; Mauche 2002)
and provide  clues to understanding the nature of  QPO phenomena.
Apparently, a common QPO feature for a wide class of CV  and LMXB
systems seems to hint  that the mechanism to produce the low-high
frequency correlation has  no strong  direct dependence
 on  the  specific   parameters of sources,
such as the  mass, radius, spin,  presence or absence of
 a hard surface of compact object and  magnetic field (see, e.g.,
Mauche 2002; TW02; Warner \& Woudt 2004ab).

  In conclusion,
 the extension in frequency for  five orders of magnitude
 in  $\nlow$-$\nhigh$ relation among  CV and LMXB  systems  may  imply
 that  the frequencies $\nlow$ and $\nhigh$ are  intimately
 related  to the parameter A = $(m/R_{6}^{3})^{1/2}$
 or the Keplerian frequency $\nk$ where the parameter A
appears (see Eq.(\ref{nt})) because the typical values of masses and
radii of WDs and NSs  can arise the variation in the parameter A for
 five orders of magnitude.
 However, clearly seen in our model,  $\nlow$ and $\nhigh$
  satisfy the above condition  and  the ratio   between them  is
   almost a constant that is nothing to do with the parameter A.
 Furthermore, for the different  CV and LMXB sources
 the variation of  $\nlow$ or $\nhigh$  is  modulated by
 the variation of parameter A,   and then  for the same source
 it is modulated by the position parameter X which  depends
  on the accretion rate. This fact
  can  explain why the QPO data distribution of CV and LMXB  sources
 in $\nlow$-$\nhigh$
 plot is homogeneous for the different sources and for the
 different accretion rates of the same source (see figure 1).
 By the way,  the QPO similarity phenomena are not  special
for the CV and LMXB  sources, which   have also been found in
  micro-quasars and quasars with 3:2 ratio in their twin-peaked QPO
   frequencies (see, e.g., Abramowicz et al 2004), and they may
  reflect the fundamental characteristics of
 accretion  flow  around any astrophysical  objects in universe.

 {\bf
 However, the $\nlow$-$\nhigh$ correlation of black-hole timing features
 are in between those of WD and NS (See Fig. 1).  To interpret this,
  we mention  the fact that for black-hole systems (as well as the
  low-luminosity neutron star-systems),  $\nhigh$  corresponds to a
 very broad feature rather than a narrow QPO (BPK02).
Moreover, to successfully apply our  model  to BHC,
which has a different environment
 from those of WD and NS,   such as the hard surface and magnetic field
 in WD or NS,
  we are inclined  to assume that a dipole magnetic field  is rooted in
  the accretion
 disk of BH and origins  at its innermost stable circular orbit (ISCO).
  So, by conveniently choices
  of the parameters like what have been done for WD and NS,
   the special radii can  be constructed
   to account for the detected relation between  $\nlow$ and $\nhigh$.
 }

Finally,  we stress  that the low and high QPO behaviors in CV and
LMXB  sources
 are so complex that no simple explanations have been likely to be adequate
(see, e.g., van der Klis 2006). As declaimed by Warner and Woudt
(2004ab), the existence of magnetically controlled accretion is in
principle testable by the further QPO detections.
 Upon  the QPO correlation in  CV and LMXB   sources,
 it essentially implies that   the QPO phenomena  generate  in the
accretion disc around any low-magnetic field compact objects, and
therefore our model would  provide a competitive choice.

\begin{acknowledgements}

We thank T. Belloni, M. M\'endez, D. Psaltis, C.W. Mauche, B. Warner
and P.A.  Woudt for kindly providing the QPO data.
Discussions with  J. Petri, P. Rebusco, J. Hor$\acute{a}$k,  V.
Karas,  G. Hasinger,   T.P. Li, J. L. Qu, L.M. Song  and S.N. Zhang
are highly appreciated.
This research has been  supported by the  innovative project of CAS
of   China.
 The authors are grateful to the anonymous referee for the critical and
 valuable suggestions for the revision of the paper.

\end{acknowledgements}

\clearpage
\begin{figure}
\includegraphics[width=11.1cm,height=18.1cm,angle=270]{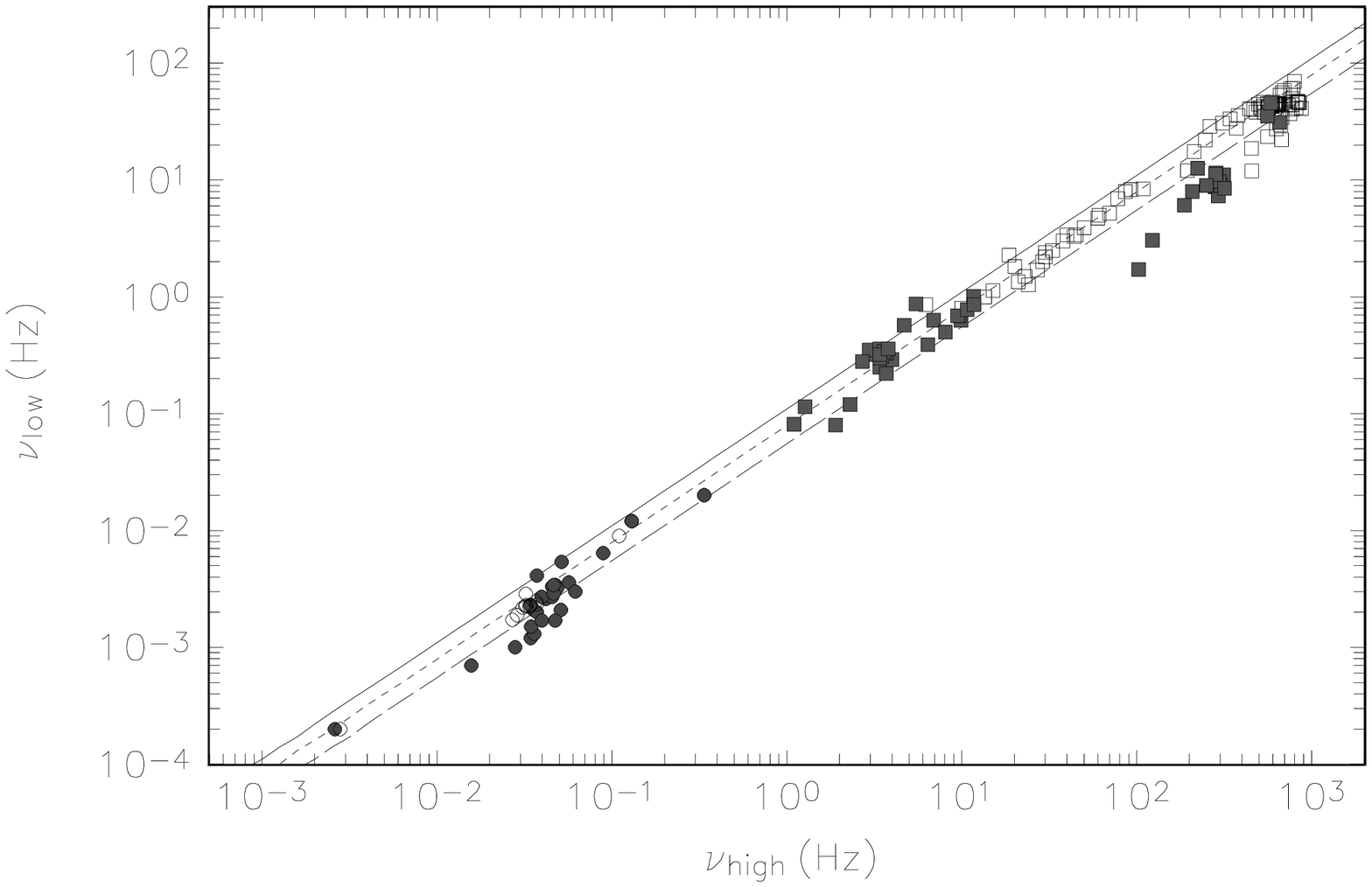}
\vskip 3.9cm
 \caption{The low QPO frequency versus high frequency
diagram for the LMXBs (filled squares: black hole binaries, open
squares: neutron star binaries), 27 CVs (filled circles) plotted
only once for each CV, and  SS Cyg and VW Hyi (open circles).
 The  QPO data of LMXBs are from Belloni, Psaltis \& van der Klis
(2002) and  kindly provided by T.~Belloni. The CV data are from
Warner \& Woudt ~(2004ab) and Mauche (2002), which are kindly
provided by B.~Warner \& P.~Woudt and C.~Mauche.
 The three straight lines represent the theoretical
curves $\nlow/\nhigh$ = 0.11, 0.08, 0.06, respectively, from top to
bottom.} \label{fig1}
\end{figure}


\end{document}